# Yury A. Lebedev[1,2,4], Pavel R. Amnuel[2], Anna Ya. Dulphan[3]

[1] Bauman Moscow State Technological University, Moscow, Russia.
[2] International Center for Everettic Studies, Israel.
[3] The National Technical University "Kharkiv Polytechnic Institute", Kharkov, Ukraine.
[4]e-mail: lebedev@bmstu.ru

# The Everett's Axiom of Parallelism

**Abstract:** In this work we consider the meaningfulness of the concept "parallel worlds". To that extent we propose the model of the infinite-dimensionaly multievent space, generating everettics altervers in each point of Minkowski's space time. Our research reveals fractal character of such alterverse. It was also found that in Minkowski's space $x, ict$ the past actively influences the present, whereas the future is a conservative factor – it slows down already occuring processes and interferes with actualization of the latent ones. Fast fusions formation is predicted based on modeling of fractal dynamics of. It was also found that the alterverse branches grow in non-Markov fashion; some of this feature are discussed. The concept "fractal parallelism according to Everett" is proposed. Inevitable inaccuracy of the model is also discussed.

> *Context of interpretations of quantum mechanics is pluralistic, as a result, notoriously abundant, but unsuccessful attempts to find the one and only "true" interpretation seem to have led by now to realization that this effort is as utopian as perpetuum mobile.*
> *Plurality of interpretations of quantum mechanics is as inevitable as the strangeness of the world that quantum mechanics discovered (or created).*
>
> V.I. Arschinov [1]

Among the dozens of interpretations of quantum mechanics seriously discussed by physicists and philosophers in recent years, two are the most significant and drawing most attention: Copenhagen interpretation and the many-worlds one. In philosophy the many-worlds interpretation is presented the form of everettic: axiomatic ideological construction, whose axioms include the most important point of the many-worlds interpretation, specifically, branching of the wave function during the interaction process [2, 3, 4].

Concepts of the many-world (everettic, as we call it hereafter) branching and fusions are basic axiomatic concepts of everettics [2]. However, Hew Everett's paper [5] does not detail the mechanism of branching, which certainly strengthened the concept of "parallel worlds" along with the respective term. That is particularity true for popular presentations of many-worlds interpretation of quantum mechanics.

"Geometric" understanding of the "parallel worlds" concept has in its core a statement about "disjointness" of alterverse[1] branches. The concept has in its basis the passage from Everett's work: " This total lack of effect of one branch on another also implies that no observer will ever be aware of any "splitting" process " [5]. As a result of interpretation of the concept of the branch in terms of epistemological optimism, everettics put forward the idea of branch fusions [6, pp. 106-107] and the postulate of "disjointness" was replaced by another one: "Axiom of everettical fusions", which proclaims the inevitable interaction between the alterverse's branches [2, p.56].

Additionally, everettics postulates Fifth axiom about metasystem of the universes. This axiom reflects the current most common conception of the structure of being: "Being as a whole, is a Godel's fractal metasystem of universes and their inhabitants" [2, p. 56].

The present work is an attempt to specify the manifestations of everettic axioms based on fractal model of the mechanism of the everettic branches formation.

Let us consider a structure of alterverse of an object A in the Minkowski event space. The question of the general physical interpretation of the event for object "A" is separate everettic issue that requires special attention.

For the purpose of this work it is important to consider an event which has universal character and clear physical meaning. In that regard, the event should be generated by the environment that is present at any point in Minkowski space-time. Physical vacuum is a logical choice in this case. From a philosophical point of view, we can consider any other model of "the aether " in its Einstein's interpretation as a filling of the void [8]. However, the model of the physical vacuum is preferred because inevitable quantum fluctuations of the physical fields in this environment play an important role in explaining some of the fundamental phenomena, not only the "exotic" ones(chaotic inflation by Linde, Hawking radiation, the Lamb shift, van der Waals forces, etc.), but "every day life" ones as well(spontaneous emission of excited atoms).

Thus, we assume that the object A is a light bulb, which is located in the cabin of a spaceship, and the event is a "flash of light" produced by this bulb. We will leave aside the technical details of the observation of this event, as well as feasibility of this observation.

---

[1] Alterverse is a set of classical realities of the physical world (CRPW), reflecting the state of the single quantum reality (SQR). The alterverse is structured in the branches as specific CRPW that are relative states of Mensky's crystal faces and consciousness of the observer. The term reflects the fact that different "Everett worlds" are different alternative "projections" of the quantum world (SQR) on the memory of the observer. The term was proposed by Mensky [7].

Let us also assume that the spaceship can move at any sub-light speed. This means that the light in the cabin is to be located in any point of future light cone of Minkowski's event space of the ship.

It is known that the each event of the photon emission by the filament of an incandescent bulb is due to fluctuation of the electromagnetic vacuum. (In the absence of such fluctuations, the excited state of the atom would be stable, and the bulb would not emit light.)

The substantiation of this event at the given point of Minkowski space is determined by the presence of a set of excited atoms (filament) and a random value of the energy of vacuum fluctuations of the electromagnetic field at that point. Point $\{x_1, y_1, z_1, ict_1\}$ where the event 1 occurred in alterverse of the object A is a branching point: the object A goes into a state that can produce flashes of light in some other points $k$ $\{x_k, y_k, z_k, ic(t_1 + \Delta t_k)\}$. Coordinates $x_k, y_k, z_k$ depend on the specific route chosen by the crew, or other reasons influencing the speed and direction of the lamp location point, and the coordinate $t_1 + \Delta t_k$ depends on an arbitrarily chosen interval $\Delta t_k$ and random vacuum fluctuations at $\{x_k, y_k, z_k, ic(t_1 + \Delta t_k)\}$. If the intensity of the fluctuations at this point is below a certain threshold, the flash of light does not occur. Therefore, the event $k$ only occurs at certain points of Minkowski space. The points $\{x_k, y_k, z_k, ic(t_1 + \Delta t_k)\}$ at which event $k$ may occur we will call active branching poins.

The axiom of everettic branchings dictates that cross section of the space-time structure of the object A alterverse by isotemporal surface $ic(t_1 + \Delta t_{1k})$ should contain the active branching points.

Not reducing generality of the model, we extend the analysis to the case of two dimensional Minkowski space $\{x, ict\}$ (Fig. 1).

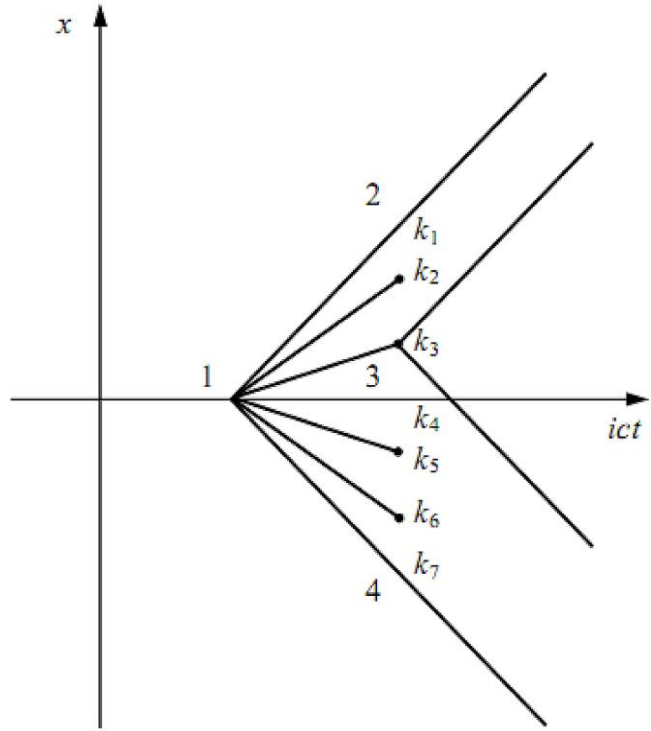

Fig.1. The object A alterverse in two dimensional Minkovsky space.

Fig. 1 showes *l* and *k* events in two dimensional Minkowski space. The bulb located inside A is on at point 1. After that, the object can move along different trajectories $1 \rightarrow k_1$, $1 \rightarrow k_2$, ..., $1 \rightarrow k_7$ during time interval $\Delta t_k$, which corresponds to change of coordinate *ict* by segment (1-3). Fig. 1 presents the case where each specific alterverse branch (direction and speed of object A movement from point 1) is chosen by the ship crew or is a result of deterministic laws of mechanics. Thus, the shown structure of alterverse branches is a macroscopic deterministic part of its overall structure, and does not reflect the branches arising from the quantum fluctuations of the electromagnetic vacuum.

Rays (1-2) and (1-4) limit the light cone of event 1. Isotemporal surface of the section of alterverse represented by the segment (2-4). The points $k_1, ..., k_7$ can potentially contain "event of flash". For clarity, it is assumed that this happened at point $k_3$, which in this case is the active branching point. This is reflected by the construction of the light cone of the event $k_3$.

Let us consider a region of space-time to the right of the surface (2-4), i.e. future of the elements of the surface. In the viscinity of $k_3$ we select a thin layer with thickness $\Delta ict$, adjacent to the isotemporal surface with $t_0$ coordinate. Obviously, on the segment of isotemporal secant (2-4) in the viscinity of $k_3$ there will be other points in which the fluctuations of the electromagnetic vacuum are intense enough to cause a flash of light. We denote them as $k_{3i}(i = а, в, с...)$. It is also obvious that these points are randomly distributed on the the segment (2-4).

We split the layer *($\Delta ict$)* in squares with a side of the axis of time *($\Delta ict$)$_j$* equivalent to the threshold fluctuation energy causing the flash (calculated from the uncertainty relation for energy and time), and the side along the spatial axis *X* equal to the linear size of the fluctuation (Fig. 2).

We now need to answer the question whether the structure of fluctuations (distribution of fluctuation energies in networks of cell built on the segment *($\Delta ict$)* in the chosen field of the future of A object) is static or dynamic?

If considered a Minkowski space {*x*, *ict*} was purely geometric, like the Euclidean (or any other metric space, which metric does not have time), the answer would be unambiguous: the parameters of fluctuations must be static.

However, the event spaces feature some properties fundamentally different from those of geometrical spaces.

Note that using the mathematical methods of the event spaces one usually does not discus or acknoledge presence of the External Observer associated with these spaces. This metaphysical object arises in everettics when analyzing the very statement of the problem of describing the universe as an isolated system. "The need for such a *special* External Observer logically inevitable, and results from the text of the Everett's article, the author and the audience who consider Everett 'isolated system' from outside are such observers" [9, p 64].

Presence of External Observer is even more evident in the event space - supratemporal analysis of mathematical and physical properties and phenomena of event spaces with *temporal* coordinate is performed from his perspective.

However, External Observer, always present in the description of the realities of event space, is not introduced into this model from outside, although it is in line with the Amakko principle: "For the sake of completeness one must multiply as much as possible the substances logically compatible with the fact considered " [10]. Here authors just highlight the presence of an External Observer in *all* models of event space, including the Minkowski space-time. The only Amakko property, which we assign to the External Observer in our model, is its ability to capture the locations of flashes of light and to store in memory the their time sequence.

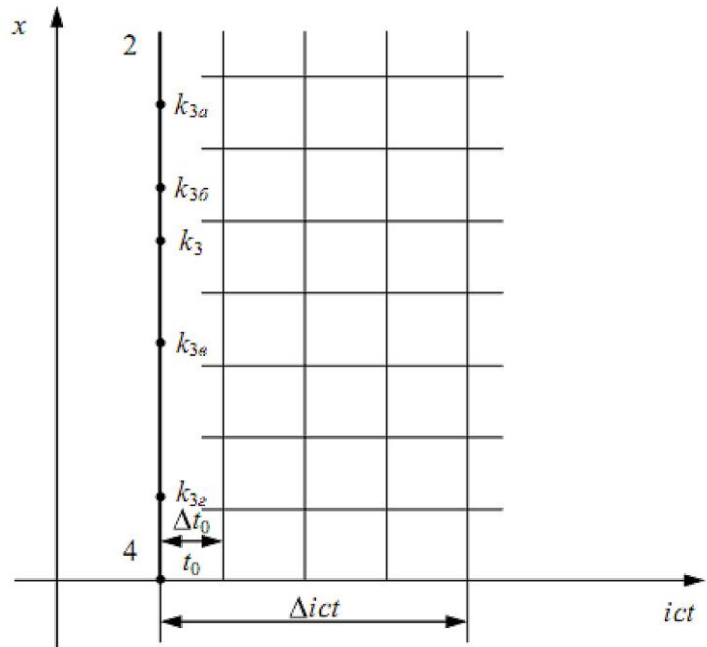

Fig 2. The area near future of A object.

One also needs to take into account the properties of the coordinates, specifically time coordinate: in the Minkowski event space it is physically impossible to capture a point $t_0$. The concept of "a moment of zero duration" (i.e. "time point") does not exist. Temporal point is defined with a precision $\Delta t_0$, and the its value depends on the accuracy of measurement of the energy of the event, in accordance with the Heisenberg uncertainty principle . Wallace defined the essence of temporal coordinate in event spaces as follows: "We may speak of 'moments of time' and the number of moments of time ('the next moment', etc. ) but this is just a metaphor for temporal duration, and cannot be interpreted literally. " [11]. Therefore the discretization performed earlier uses a lattice, where the parameters of vacuum fluctuations in each cell are random variables, determined by the physical properties of the vacuum in the area of the partition.

The specific value of this parameter is determined by the "here-and-now-for-me" principle. In other words, the values of fluctuation parameters at each $\{x, ict\}$ of event space will be different for different observers, or for the different calls to this point made by the same External Observer. This property of event space can be described phenomenologically by the notion of "intrinsic time $\tau$" at each point of event space. Mathematically, this is equivalent to introduction of one more dimension at each point {*x, ict*}, orthogonal to both *x* and *ict*.

This dimension should have characteristic of the time (in this case, the most important characteristic is fluidity) and have dimensionality of $ic_\tau\tau_\tau$. We leave aside the issue of the the value of the constant. Thus formed space {*x, ict, $ic_\tau\tau_\tau$*} is infinite multi-event space, and its corresponding section {*x, $ic_\tau\tau_\tau$*} at *ict = const* corresponds to everettic alterverse of events at $k_3$. This approach, as opposed to approach of External Observer, is a direct consequence of the Amacco principle applied to this system. Moreover, in this case the Amacco principle is used in its strictest form - the model considered has an infinite number of new entities.

The space is essentially a universal state of object A space. According to Wallace: " We are undoubtedly more at home with Minkowski spacetime than with the universal state. Partly this may be because we have worked with the concept in physics for rather longer, but more importantly we have long been used to the idea that multiple times exist (in some sense) — the innovation in relativity theory is the unification of these instants into a whole, and the identification of the instants as secondary concepts. Everett asks us to take both steps at once: to accept that there exist many worlds, and then to fuse them together into a whole and accept that the worlds are only secondary. ". [11]

An important feature of the space {*x, ict, $ic_\tau\tau_\tau$* } is the fact that there is no single point of "origin" - each event has its alterverse, i.e. $ic_\tau\tau_\tau$ axis occurs at each point of axis *ict*.

Introduction of the alterversal space {*x, $ic_\tau\tau_\tau$* } allows us to move on with alterverse of the flash of light on the object A in the viscinity of $k_3$ in the Minkowski event space.

To proceed further, It is important to understand a certain feature of uncertainty relation for energy and time:

$\Delta E \Delta t \geq \hbar$

Applying this relationship to the point $t_0$ (Fig. 2), one can see two potential outcomes of the energy fluctuation :

First: $\Delta t > 0$, and $\Delta E > 0$. That means that at $\Delta ic(t_0 + \Delta t_{10})$ (i.e. in the future of the point $t_0$), the energy of the lattice element to the right of $t_0$ is greater that the energy of $t_0$. Based on the principle of local energy conservation this fluctuation means less energy in present time and increase it in the future.

Second: $\Delta t < 0$, and $\Delta E < 0$. This means that at $\Delta ic(t_0 - \Delta t_{10})$ (i.e. in the past of the point $t_0$), the energy of the lattice element to the left of point $t_0$ is less than the enrgy of $t_0$. Based on the principle of local energy conservation this fluctuation means more energy in the present time and decrease in the past.

We now see that in the space {*x, ict*} only the past actievely influences the present (adding energy stimulates actualization of latent processes), while the future conservatively influences the present (energy decrease slows down already occuring processes and hinders actualization of the latent ones).

However, External Observer in the space {*x, ict, $ic_\tau\tau_\tau$*} will see it differently. In a supratemporal plane {*x, ict*} selected by the External Observer in the absence of object A, fluctuation of energy in every cell of the lattice will randomly vary over time $\tau_j$ in alterverse spaces $\{x, ic_\tau(\tau_j)\tau_j\}$. External Observer will therefore capture a picture of the cells that contain the energy necessary for flash of light at the point $k_3$, which will correspond to the equilibrium Brownian motion of points (cells with threshold energy sufficient for the flash) on the part of the plane $\{x, ict\}$ within the light cone of point $k_3$. Fig. 3 shows possible displacements of one of the observed elements of the "apparent perturbation" along the mesh of elements of alterverse spaces.

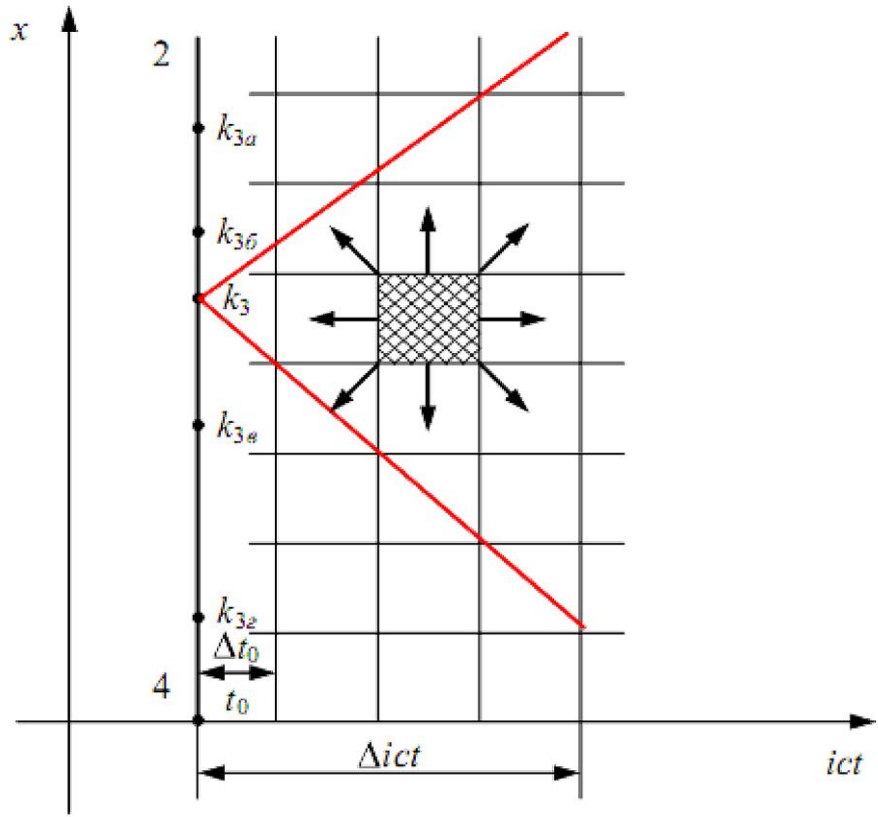

Fig. 3. Displacements of "apparent perturbation" in network of alterverse cells.

When the A object appears on line (2-4), the physical conditions of that line change: a "scavenger fluctuations" arises at the point $k_3$ – an excited atom in the filament of lamp. A similar pattern can be observed for all points $k_{3i}$.

In this case Brownian motion of the points of the "effective disturbance" will transform, according to Le Chatelier-Brown principle, into the diffusive motion towards excited atoms.

Considering that a real flash in physical space (which in this case is represented by a plane $\{x, ict\}$) occurs during a finite period of time $\Delta t_0$, "flowing" along the axis $ict$, and can occur in any cell adjacent to the cell containing the point $k_3$, one can see that the sequence of flashes (alterverse event at point $k_3$) will look to External Observer as a dendrite growing from point $k_3$. As shown in [12], the type of fractal of branching in this case will depend on the conditions of the structure formation.

To describe this process, we applied the model of a random fractal developed by A. Dulfan in his work "The random fractal with a given preferential direction of growth" [13]. The model is based on the Witten-Sander method.

The method is based on a concept of a fluctuation randomly occuring in the lattice And then stochastically moving up until it "encounters" the element, which the External Observer captures. The algorithm of the simulation is detailed in [12].

Numerical simulation performed in [12], which we interpret in terms of our alterverse model, shows the pattern of growth of fractals of alterverse branching of events at $k_3$, occuring at various local conditions for the origin and motion of fluctuations.

Fig. 4 showes a graphical representation of the simulation results for a single "active spot".

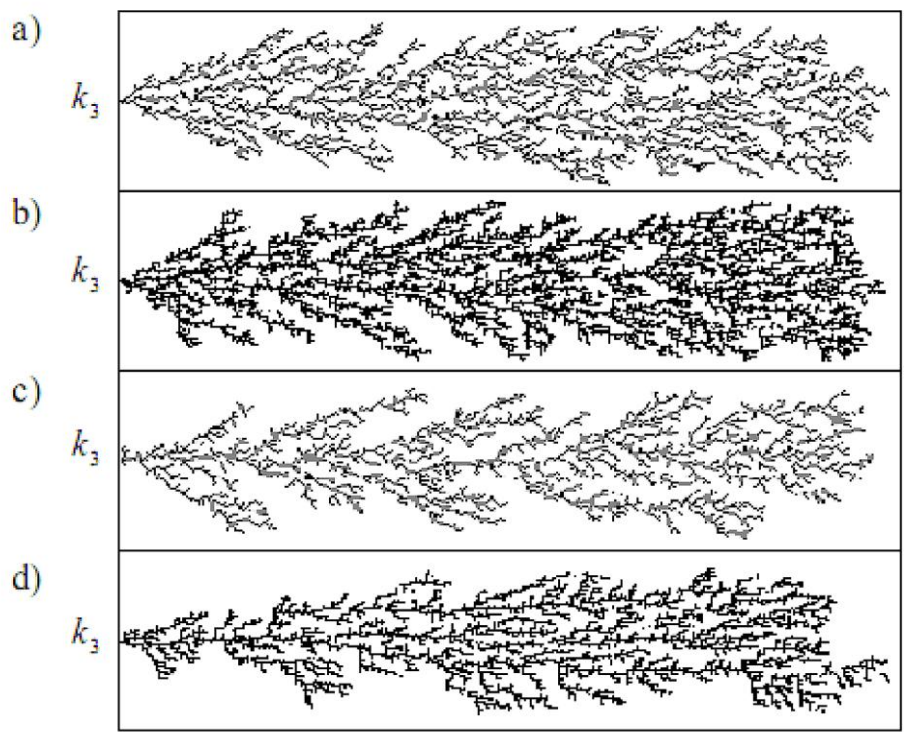

Fig. 4. Fractal growth of the various points in the diffusion mode.

The monostructures of alterverse resulting during the generation of fluctuations at various degree of isotropy are shown in Fig. 4:

a) isotropic situation: movement of fluctuations is only possible in the horizontal and vertical directions, "fluctuations absorbtion" occurs on the same lines,

b) partially isotropic situation 1: fluctuations only move in the horizontal and vertical directions, and absorption is possible in both of these and the diagonal direction as well,

c) partially isotropic situation 2: fluctuations can move not only in the horizontal and vertical directions, but also along the diagonals; absorption is possible only in the horizontal and vertical directions,

d) anisotropic situation: fluctuations move horizontally, vertically and diagonally; absorption of fluctuations takes place in all these directions as well.

Fig. 4 demonstrates that the alterverse fractal depends only weakly on the diffusion and steric factors (direction of interaction between the excited atom and fluctuation), which reveals the stability of the model in the presence of heterogeneity of local conditions.

This us gives reason to consider growth of alterverse from several points $k_{3i}$.

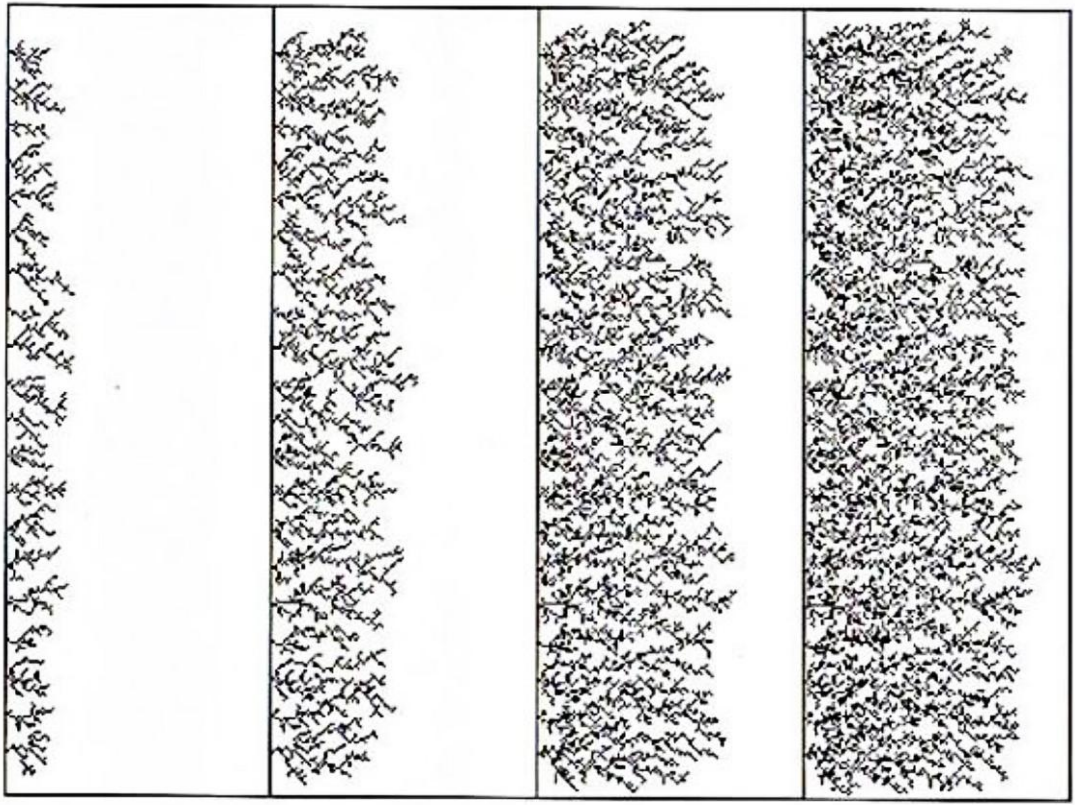

Fig. 5. Dynamics of the alterverse branching.

Fig. 5 shows four of the 9000 sequential steps of the modeling (Witten-Sander method, at the increment $\Delta t_0$) of the location of the "apparent fluctuation". Note that the term "dynamics" in our alterverse model has a specific meaning. The pattern of events represented by dots in Fig. 5, is not directly related to the dynamics of flashes in the event space $\{x, ict\}$. Rather, it is the "road map" of one of the layers of space $\{x, ict, ic_\tau \tau_j\}$, captured by External Observer.

Its physical meaning is that it predicts flashes of light at certain points of the segment (2-4) at $n\Delta t_0$ time intervals in event space $\{x, ict\}$ under the condition of frozen times $\tau_j$ (Isochronous section of space $ic\tau_\tau$). (Fig. 6)

An obvious feature of this fractal structure is the large number of alterverse branches intersections, considered to be realities fusions s in everettics.

An important detail emerged from consideration of a detailed modeling of the alterverse evolution is the fact that branch fusions occur already at a relatively small number of steps. Thus, in Fig. 6 alterverse branches of $k_{3b}$ and $k_3$ intersect at step 19, and the alterverse branches of $k_{3c}$ and $k_3$ intersect at step 13.

This same pattern was found for branches of any $n$'s section of alterverse if n is sufficinly large.

One can guess that this feature is characteristic of most other fractal models of everettic branches.

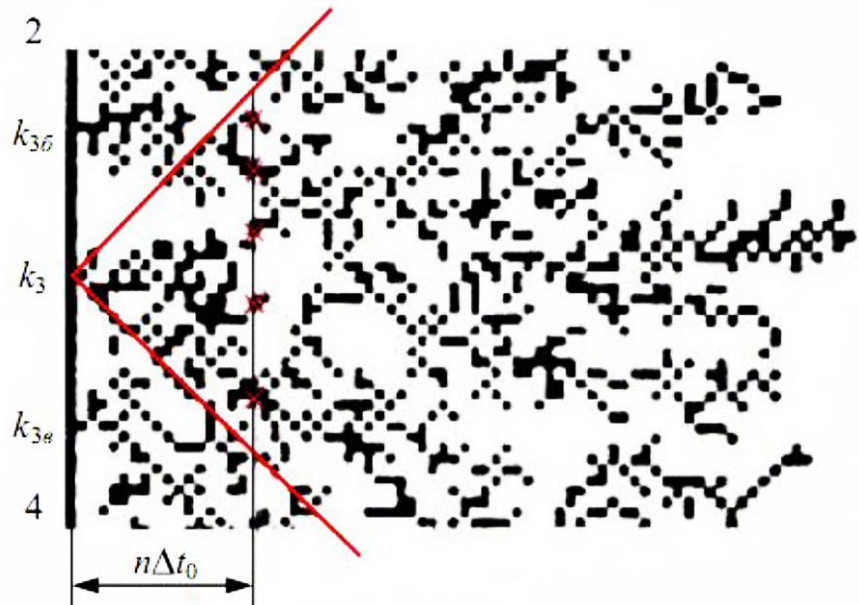

Fig. 6. Spots of light flashes (red crosses) on the surface (2-4) within the light cone of the point $k_3$ at intervals $n\Delta t_0$ in space $\{x, ict\}$.

Complete the "road map" for a future of $k_3$ should be in the n-dimentional space and be a dynamic object in each of $n$ times $\tau_j$ of alterversal spaces $\{x, ic_\tau(\tau_j)\tau_j\}$. Moreover, analysis of Heisenberg's uncertainty principle $\Delta E \Delta t \geq \hbar$ revealed that $k_3$ and the object A should have a similar structure of their "road map" of the past.

Since in alterverses of the times $\tau_j$ both "road maps" are dynamic objects that have a common point, there is no reason to dismiss their interaction and mutual influence. Moreover, for the half cone of the future, the fractal considered "is essentially a non-Markovian and therefore it is very difficult to study analytically " [12]. This means that not only deterministic, but also random events in space $\{x, ict\}$ depend on the evolution of the system as a whole. (In our case, the former are the ones of object A appearing at points $k_i$ (Figure 1), which are due to the decisions of the crew, and the later are the random events of the flashes of light at $k_{3i}$ in Figures 2-5).

Non-Markovian character of the evolution of alterverse branches allows us to resolve persistent questions regarding the description of the features of certain quantum paradoxes. For instance, the following problem is posed by the famous paradox of Schrödinger cat.

In a closed box Schrodinger cat exists in a superposition of its possible states. Let us assume that , after the is opened box, we find a live cat. This would mean that a dead cat was in the other multiverse branch. Close the box again, and wait for a while, then open the box. Suppose that we again see the cat alive. So, there arises another alterverse branch with a dead cat. Let us now repeat this procedure until we finally find a dead cat. Now we are in a branch of a dead cat, and the number of such branches is $N$. With reagrd to the cat all these branches are the same - cat is dead in all of them.

What is different in each branch is the external event: in one branch the technician caught a flue, in another he had dinner and so on. However, it is never mentioned in the procedure description, and the fate of the obsever during the experiment is normally omitted from consideration. Non-Markovian nature of everettics branches predicts that the presence the information about the death of a cat in the memory of the observer limits his subsequent behavior and, therefore, structures his future. For instance, in those alterverse branches where a cat died, technician will never come to the experimental box with a bowl of milk. It is however very likely in the branches, where cat was alive in the preceding opening of the box.

This means that the entropy of the future of non-Markov processes in the alterverse (processes, depending on the history and memory of the observer) is always less than the entropy of the future of Markov processes that are independent of history. For a more detailed discussion of the entropy in the alterverse evolution an improved algorithm of fractal simulation is needed, one accounting for the memory of External Observer.

Due to its symmetry, fractal of alterverse past for the object A and for the point $k_3$ is non-Markov, and the entire space $\{x, ict\}$ is "historically conditioned" regardless of the origin and the direction of the axis $ict$.

Obviously, the scales of the axes $X$ and $ict$ on Figs. 1 and 2-6 differ by many tens of orders. Moreover, the volumes of the event spaces of the ship (object A) and nano-sized element of the lamp filament in its cabin, containing the point $k_3$ (hundreds of orders for the four dimensional Minkowski spacetime) differ as well.

Once we realize that is practically impossible to build a "road map" of alterverse of past and future for the point $k_3$ at the current computational level, calculations of these cards for both micro-and macro-objects becomes seemingly hopeless.

However, the many-worlds interpretation of quantum mechanics, being a part of the ideological foundation for quantum computers, may obtain a tool for quantification of alterversal spaces as a result of the development of such computers.

The proposed model offers a new perspective of the "parallel worlds". The key property of fractal is scale invariance or, in other words, a complete $\{x, ict\}$ self-similarity of the geometrical descriptions of the fractal process. Fractal in event spaces in our model adequately describes the physical processes of galactic to the atomic scale, and it can be perceived as a kind of "parallelism". However, this parallelism is not linear, as in Euclid geometry, but fractal. Note that there is not the term "parallel" in fifth Euclid postulate: "5. That, if a straight line falling on two straight lines makes the interior angles on the same side less than two right angles, the two straight lines, if produced indefinitely, meet on that side on which are the angles less than the two right angles." [14]. The property of the lines described by Euclid is only geometric meaning of the term in its current understanding. Currently, the term "parallel" is defined as "the same, a comparable" [15, page 516], which is very close to the meaning of the term "fractal".

From this perspective, fifth everettic axiom, as well as Euclid's fifth postulate, may be regarded as "an axiom of parallelism." As such, it deserves the name of the Everett's axiom of parallelism.

The meaning of the Everett's remark, cited at the beginning of this article, does not imply the absence of the splitting process. Knowledge about branching is a characteristic of the observer, not the process.

Everett's remark was sagacious in the sencese that observer taking part in the process of branching (such as those associated with the point $k_3$ in Fig. 6), loses its primary identity in a few "steps of branching" (in our example, 13 and 19 steps) and becomes the new "mixed observer" $k_{3c} - k_3$, or $k_{3b} - k_3$ mixing and loosing his initial identity progressievely. Therefore, "the initial observer" in fact ceases to exist after the first fusions and "does not know about any process of "splitting "".

The fractal nature of the Everettic "parallelism" reconciles us with the common term "parallel worlds", assuming generilized interpretation of parallelism.

To describe the everettic branching of alterverse in event space, one can use its dimension $\alpha$. In this case the "branching factor" is one temporal coordinate, therefore events with only one outcome will be characterized by an integer dimension equal to unity. The presence of branching increases the value of $\alpha$ proportionally to the density of the branches in the event space. This density limit (if there is branching at every point) would be the value $\alpha = 2$. Thus, everettic branching in the two dimensional event-space should be in the range $1 \leq \alpha \leq 2$.

Clearly, in the n-dimensional space the relation is $1 \leq \alpha \leq n$. In the case of $\alpha = n$ branching occurs at every point of event space and alterverse takes up all cells in Fig. 2 (in the model case in Fig. 5, 6, $\alpha = 1,3$).

For a continuous space the equality $\alpha = n$ means infinite number of branches and the density of the number of branches. Thus, the considered fractal model confirms utility of the modeling of Minkowski space-time by discrete networks such as Fig. 2.

It can be assumed that the fractal dimension of time keeps the information, determining the hierarchical structure of event space.

In conclusion we would like to note, that the of spontaneous radiation in space $\{x, ict, ic_\tau \tau_\tau\}$ is not quite correct example, as "point of the flash" are not captured in the cross-section $\{x, ict\}$. This capturing is only possible in the supratemporal consciousness of External Observer. The authors are aware that the granting of an External Observer the ability of such capturing may be a mistake.

Moreover, the magnitude of the interval $\Delta t_0$ in the model should be of the order of Planck time ($\sim 10^{-43}$ s) in order to ensure the applicability of the assumed model of diffusive motion. For larger intervals relativistic limitations on the vertical movement of fluctuation will take effect.

Therefore, our model is only a first approximation of fractal description of alterverse with its inevitable coarsening and inaccuracies. We hope that its further development will identify the " diffusion modes" in which conditions of Conway-Cohen theorem are satisfied[16].

However, the authors firmly believe that the "trial and error, the usual method of investigations in science, requires the consideration of all kinds of ideas, of which only one will be correct and will remain for the future" [17].

Trial and error method can be likened to a collapse of the wave function. Any scientific research (and not only scientific) is similar to quantum event, which can develop in many ways, but at the time when we observe it we see only one option, and all the other solutions of the wave equation collapse (in the Copenhagen interpretation).

However, many-worlds interpretation of scientific research is also possible: all of our research leads to the goal, but each goal is achieved in its own universe, where the laws of physics correspond precisely to such a solution. In this case the trial and error method is akin to a particular solution of "the wave equation of knowledge", randomly selected from the whole set of solutions, because the trial and error method does not investigate all possible options.

Zwicky morphological method [18, 19] allows us to consider all possible options for research, that is, by analogy, comparable to a full solution of the wave equation. Then "Zwicky morphological box" is similar to everettic multiverse: all cells of this "box" correspond to solutions of a certain problem, but every decision is executed in its own universe. Our universe corresponds to one of the cells of the morphological box.

High dimensionality of the morphological box precludes practical applications of the Zwicky method to real everettical problems. "A box composed by Zwicky to predict only one type of rocket engines, had - with 11 axes - 36,864 combinations!.." [20, p. 53]. But, as noted earlier with respect to the calculation of the "road maps of alterverse", everettics itself can evolve into a tool for the quantitative description of highly complex tasks.

The axiom of parallelism for Everett is one of the steps of this development.